# Exploring the production of Terbium-161 in the Brazilian Multipurpose Reactor

Fernando C. Melges[1,2,*], Jhonatha R. Santos[1], Luiz P. de Oliveira[1], Alexandre P.S. Souza[1,2], Carlos G.S. Santos[1,2], Iberê R.S. Júnior[1], Barbara P.G. Silva[1], Marco A.S. Pereira[1], Frederico A. Genezini[2]

[1]Reator Multipropósito Brasileiro – RMB/CNEN, 2242 Prof. Lineu Prestes Av., São Paulo-SP, Brazil;
[2]Instituto de Pesquisas Energéticas e Nucleares – IPEN/CNEN, 2242 Prof. Lineu Prestes Av., São Paulo-SP, Brazil

## ABSTRACT

The Brazilian Multipurpose Reactor (RMB) was conceived to meet national needs for radioisotope production, materials irradiation testing, and neutron beam applications. In addition to its 30 MW pool-type reactor, the RMB complex will include additional facilities for radioisotope production, and related applications. Tb-161 is a promising radionuclide for radiopharmaceutical therapy, offering decay properties similar to Lu-177 but with additional conversion and Auger electrons that enhance dose delivery to cancer cells. In light of this emerging radioisotope, this study explores the potential production of Tb-161 in the RMB through neutron irradiation of enriched $Gd_2O_3$ targets. Monte Carlo (MCNP) simulations provided detailed neutron flux distributions, which were used as input for ORIGEN calculations of isotope buildup. Assuming 10 mg targets enriched to 97.5% in Gd-160, a 40-day irradiation, and a thermal flux of $2 \cdot 10^{14}$ n/cm²s, the results indicate that Tb-161 activity reaches approximately 4.5 GBq after 14 days (≈450 GBq/g), in agreement with data from other research reactors. Building on prior studies that demonstrated the RMBs capability to irradiate larger targets, terabecquerel-scale yields appear feasible. These findings highlight the RMB's potential to support domestic production of emerging therapeutic radionuclides such as Tb-161. The potential for isotopic enrichment of Gd-160 using the AVLIS method is also discussed.

## 1.    INTRODUCTION

The radionuclide Tb-161 has attracted significant attention in recent years as a potential alternative for Lu-177 in radiopharmaceutical therapy [1-3]. Tb-161 emits β- particles with a mean energy of 154 keV and has a half-life of 6.89 days, closely resembling the decay characteristics of Lu-177. However, Tb-161 also emits conversion electron and Auger electron with higher linear energy transfer than β- electrons. This property of Tb-161 suggests that it could potentially deliver a higher dose in therapeutic applications compared to Lu-177, which has been rigorously investigated recently. Theoretical dose calculations with Monte Carlo-based codes have indicated higher dose delivered by Tb-161, compared to Lu-177 [4-5]. This theoretical advantage has been corroborated by in vitro and in vivo studies and, more recently, several clinical trials are underway and have shown promising preliminary results [6-9]. In summary, Tb-161 is expected to deliver a higher dose to cancer cells than Lu-177 due to its radiological characteristics, without causing significant adverse effects [7,9].

The irradiation of highly enriched $^{160}Gd_2O_3$ targets (enriched between 97,5% and 99,988% [10-11]) is the main route for the production of Tb-161. During the irradiation, the neutron capture of Gd-160, followed by the decay of Gd-161 (3.66 min), produce Tb-161, via the reaction $^{160}Gd(n,\gamma)^{161}Gd \rightarrow {}^{161}Tb$ [2,11-12]. The produced terbium is then separated from contaminants of the target via chromatographic purification, such as strong cation exchange resins or extraction chromatography resins [2,6,13]. This method guarantees compliance with stringent radiopharmaceutical purity requirements and supports large-scale, reproducible production under reactor conditions.

Some of the main challenges in the production of Tb-161 are related to the purity and availability of the target material ($^{160}Gd_2O_3$), as well as the purification processes required to remove contaminants and neighboring lanthanides to achieve a high specific activity product [2, 11,13]. Nevertheless, the irradiation of enriched $^{160}Gd_2O_3$ targets in research reactors, followed by chemical separation of the produced Tb, has proven to be a reliable method for obtaining high-purity Tb-161 [6,11,13-15]. Another critical challenge, however, is the limited availability of high-flux reactors, as many of those currently used for irradiation are approaching decommissioning, and the capacity of the remaining facilities is unlikely to meet the growing demand for neutron-rich radionuclides such as Tb-161 [2,13].

In this context, the present paper aims to evaluate the capability of the Brazilian Multipurpose Reactor (RMB) to produce Tb-161 via the irradiation of $^{160}Gd_2O_3$ targets. First, the isotopic enrichment of Gd is discussed, as it represents a critical step in the production process. Next, the RMB complex is briefly described, followed by an analysis of the irradiation of $^{160}Gd_2O_3$ targets in the reactor. Finally, the theoretical approach adopted to assess the reactor's potential for Tb-161 production is presented, based on the reactor's irradiation profile and post-irradiation radioisotopic inventory obtained from simulations using MCNP [16] and ORIGEN [17].

## 2.    ISOTOPIC SEPARATION PROCESS

The isotopic enrichment of Gd-160 represents a crucial and indispensable step in the reactor-based production of Terbium-161 for therapeutic use in nuclear medicine [12,18]. The enrichment process plays a fundamental role in achieving high radionuclidic purity of the final Tb-161 product and preventing the co-production of long-lived radioisotopic impurities. The most significant contaminant is Terbium-160, a radionuclide with a half-life of 72.3 days produced through neutron activation of residual Tb-159 impurities in the gadolinium target [1-2,13]. The presence of Tb-160 is undesirable, as it contributes to unnecessary radiation exposure to patients and complicates radioactive waste management. To ensure clinical-grade quality, the radionuclidic purity of Tb-161 must exceed 99.9% [12].

In addition to purity considerations, isotopic enrichment is essential to achieve the high molar activity required for radiopharmaceutical applications and to overcome the intrinsic limitations of natural gadolinium, which contains only 21.86% of the target isotope Gd-160 [11]. Other naturally abundant isotopes—such as Gd-155 (14.80%) and Gd-157 (15.65%)—exhibit extremely high neutron-capture cross sections of 60,330 and 254,000 barns, respectively, producing a self-shielding effect that significantly reduces the neutron flux available for Gd-160 activation and consequently decreases the production yield of Tb-161. By enriching the gadolinium target in Gd-161, the impact of these competing isotopes is minimized, allowing for efficient neutron utilization and the formation of Tb-161 in a no-carrier-added form [18]. Consequently, Gd-160 enrichment directly affects the radiochemical yield, molar activity, and overall therapeutic quality of Tb-161, consolidating its role as a key parameter in the development of next-generation theranostic radiopharmaceuticals [3,10-11].

The most effective enrichment technique for this isotope is Electromagnetic Isotope Separation (EMIS), whose operating principle is based on the deflection of ions in a uniform magnetic field, as described by the relation [3]

$$r = \frac{mv}{qB},\qquad(1)$$

where $r$ is the radius of the ion trajectory, m the ion mass, v its velocity, q the ionic charge, and B the magnetic field intensity. According to this relation, ions with different masses follow distinct circular paths, allowing their selective collection at specific focal points corresponding to each isotope as illustrated in Figure 1.

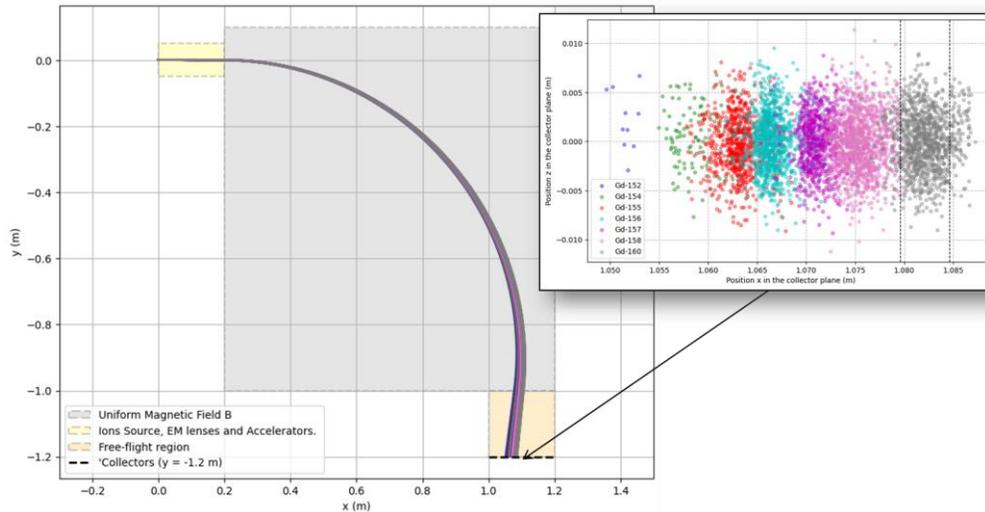

*Figure 1. Simulation of ion trajectories of gadolinium isotopes inside a 90° magnetic separator. Simulations were performed for 1 mA ion beams accelerated to 50 keV in a region with a uniform magnetic field of 0.4 T, taking into account possible optical aberrations and space-charge effects. The enlarged inset shows the spatial distribution of ions impacting the collector plane. Assuming a collection slit width of 5 mm, the resulting targets exhibit an enrichment level of 98.5% with a production rate of 1.06 mg/h.*

Although the EMIS technique provides excellent isotopic purity and mass resolution, it exhibits significant limitations in terms of productivity and energy consumption, as it operates with low-current ion beams and requires high magnetic fields and ultra-high vacuum conditions. The separation rate is further constrained by space-charge effects and the relatively low efficiency of ion sources, typically restricting production to only a few grams per year. In contrast, the Atomic Vapor Laser Isotope Separation (AVLIS) technique has emerged as a promising alternative to overcome these constraints. Based on the selective photoionization of vapor-phase atoms using lasers tuned to resonant frequencies specific to the desired isotope, AVLIS enables high isotopic selectivity, superior quantum efficiency, and modular scalability, while substantially reducing the energy demand and operational costs associated with EMIS. Studies have demonstrated that gadolinium exhibits suitable electronic transitions for selective excitation [19-20], thereby making AVLIS a viable route for the enrichment of Gd-160 with the potential for higher production yields.

In recent years, a series of studies have been carried out in Brazil with the objective of developing national expertise and achieving technological autonomy in the isotopic enrichment of rare-earth elements using the AVLIS method [21-23]. Building upon these efforts and aligned with the objectives of the Brazilian Multipurpose Reactor (RMB) program, a new phase of research has been initiated to apply this knowledge toward the production of isotopes of interest for nuclear medicine, such as Gd-160 and Yb-176.

## 3. THE NEW BRAZILIAN NUCLEAR REACTOR

Brazil currently operates four nuclear research reactors: the 5 MW IEA-R1 and the 100 W MB-01, located at the Institute of Energy and Nuclear Research (IPEN) in São Paulo; the 340 W Argonauta, located at the Institute of Nuclear Engineering (IEN) in Rio de Janeiro; and the 100 kW TRIGA IPR-01, located at the Center for Nuclear Technology Development (CDTN) in Belo Horizonte. The demand for radioisotope production for the Brazilian healthcare system, materials irradiation testing, and neutron beam applications necessitates national self-sufficiency in the development of nuclear technologies. To meet this objective, the National Nuclear Energy Commission (CNEN) has developed the Brazilian Multipurpose Reactor (RMB) project, which is under construction in the city of Iperó, 100 km from São Paulo [24]. Construction began in 2025, and the RMB is projected to achieve criticality in 2032 with a power output of 30 MW.

RMB will be a pool-type reactor, moderated by light water, and will employ heavy water and beryllium reflectors. The reactor core will be composed of 23 fuel elements containing low-enriched uranium (19.75%) with a power density of approximately 312.5 W/cm³, capable of producing a neutron flux of 4×10¹⁴ n/cm²s [25]. RMB complex is planned to include facilities for radioisotope production and quality control, analysis of irradiated material, nuclear waste treatment, buildings for researchers, an educational center with a library, mechanical workshops, among other support facilities. The unit responsible for the application of neutron beams to study the structure of matter will be the National Neutron Beam Laboratory (LFN), a facility that will feature a suite of 15 neutron instruments, including diffractometers, reflectometers, and spectrometers, in addition to support laboratories for sample preparation and pre-characterization [26].

RMB's reflector tank features 20 irradiation positions (Figure 2) for the production of radioisotopes, including Molybdenum-99 (Mo-99), Iridium-192 (Ir-192), and Iodine-131 (I-131). However, the RMB's design allows for an expansion of the types of radioisotopes that can be produced, in accordance with the reactor's planned operation and safety criteria. Irradiation position #17 has an estimated thermal neutron flux of 2·10¹⁴ n/cm²s, a value sufficient for producing radioisotopes via either nuclear fission or neutron capture processes. As investigated by Oliveira *et al.* [25], the RMB is a multipurpose reactor capable of operating at maximum radioisotope production capacity—that is, with all irradiation positions filled with targets—without compromising other facility activities. The irradiation of a Gd-157 target demonstrated the capability to irradiate targets with a high neutron absorption cross-section without overall detriment to the reactor. Therefore, the irradiation of Gd-160 targets for the production of the unstable Gd-161 is a fully viable route within the RMB's infrastructure.

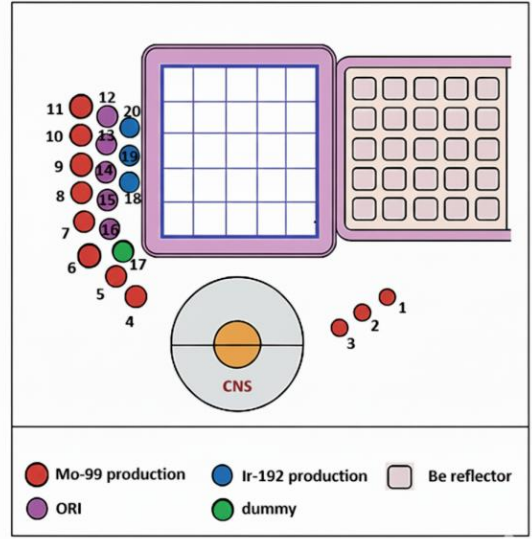

*Figure 2.Schematic diagram of the RMB irradiation positions for radioisotope production with the legend at the bottom. In this study, Tb isotope production will be tested on the dummy (shown in green).*

## 4. THEORETICAL APPROACH

ORIGEN (Oak Ridge Isotope Generation code) is a module of SCALE, a code system that includes verified and validated tools for the nuclear industry, e.g., for criticality calculations, reactor physics, shielding, and thermo-source characterization. ORIGEN [17] performs calculations of generation, transmutation, and decay through a system of differential equations, as shown in Equation (2) below:

$$\frac{dN_i}{dt} = \sum_{j \neq i} \left( l_{ij} \lambda_j + f_{ij} \sigma_j \Phi \right) N_j(t) - (\lambda_i + \sigma_i \Phi) N_i(t) + S_i(t), \tag{2}$$

where $N_i$ is the quantity of atoms of nuclide "i", $\lambda_i$ is the decay constant of nuclide "i" given in [s⁻¹], $l_{ij}$ and $f_{ij}$ are the production fractions of nuclide "i" from nuclide "j", by decay and by transmutation, respectively, $\sigma_i$ is the average removal cross-section of nuclide "i" (cross-sections collapsed to a single energy) given in barns, $\Phi$ is the average neutron flux (n·cm⁻²·s⁻¹) in the analyzed space and Si is a feed term for nuclide "i" in atoms per second, which allows the inclusion of external sources. Equation (2) has no spatial dependence and its solution must be interpreted as a spatial average over the analyzed volume [17].

In the context of Tb-161 production, ORIGEN is capable of calculating, as a function of irradiation time, the activities of all isotopes generated or depleted simultaneously through the neutron capture of enriched Gd-160 targets, transmutations, and radioactive decay. This makes it a comprehensive tool for estimating the isotopic composition of the target after irradiation.

In order to investigate such a process in RMB, it is necessary to inform the available neutron flux that will irradiate that target. To fulfill this requirement, we have used Monte Carlo simulations to obtain this input. Utilizing such MCNP code [16], we successfully generated a comprehensive neutron distribution profile for the dummy position (Figure 3), where the Gd-160 is supposed to be irradiated. Within the neutron flux distribution profile, it has been possible to produce an appropriated input for the SCALE software.

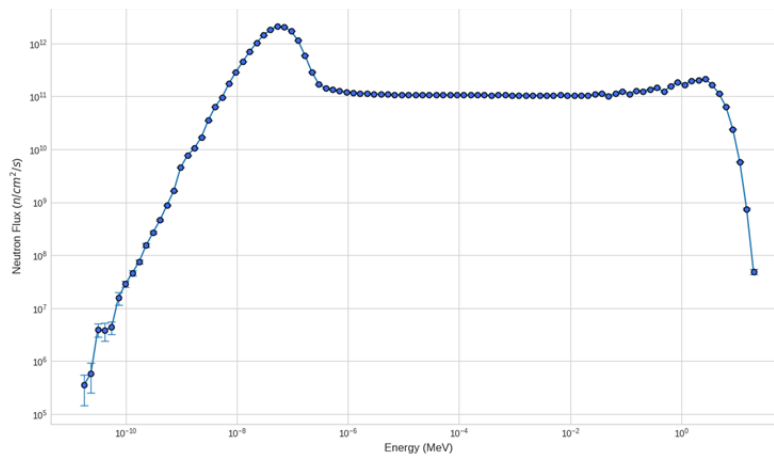

*Figure 3. Spectrum of neutrons detected at the dummy position as a function of energy.*

The fitted curve $\phi(\lambda)$, as a function of the neutron wavelength $\lambda$, for the thermal neutron region (E < 0.65 eV) has the following form and parameters, $\eta = 1.10 \cdot 10^{20}$ n/cm²s¹K²Å⁵, T = 286.70 K, $\gamma = 1102.19$ KÅ², $R^2 = 0.99772$,

$$\phi(\lambda) = \frac{\eta}{\lambda^5 T^2} \, exp \left(-\frac{\gamma}{T\lambda^2}\right). \qquad (3)$$

The parameters adopted as input for the ORIGEN simulation are presented in Table 1. No potential contaminants of the $Gd_2O_3$ targets were considered in the simulation.

*Table 1. Input data for ORIGEN.*

| Target type | $Gd_2O_3$ (97.5% Gd-160) |
|---|---|
| Target mass | 10 mg |
| Irradiation time | Up to 40 days |
| Thermal neutron flux | $2 \cdot 10^{14}$ n/cm²s |

## 5. RESULTS AND DISCUSSIONS

Figure 4 shows the estimated content of Tb-161 after the irradiation of the $Gd_2O_3$ targets for up to 40 days. The activity of Tb-161 continues to increase during the 40 days reaching a maximum of 5.8 GBq. After 14 days of irradiation, the calculated activity of Tb-161 is approximately 4.5 GBq, and the specific activity per gram of the irradiated target is 450 GBq/g, comparable to what was achieved under similar irradiation conditions on other research reactors [6, 27].

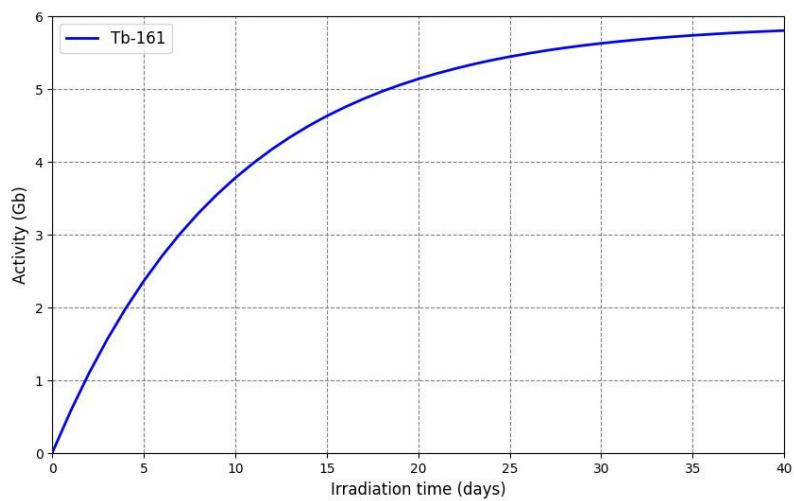

*Figure 4. Activity of Tb-161 as a function of irradiation time at the end of irradiation (EOI).*

Recent work [25] has demonstrated the irradiation potential and versatility of the Brazilian Multipurpose Reactor (RMB), indicating, based on the total neutron flux available in the various irradiation positions, that the reactor could readily accommodate higher-mass targets (>10 g) and achieve activities in the terabecquerel range for Tb-161. Naturally, several factors must be carefully assessed before implementing such high-mass targets, including the adaptation of existing radiochemical separation procedures and considerations related to the enrichment of large quantities of Gd-160. However, should the demand for Tb-161 increase to a level that justifies this scale of production, the RMB has been shown to possess the irradiation capacity to meet it.

## CONCLUSIONS

The present study discusses the capability of the Brazilian Multipurpose Reactor (RMB) to produce significant quantities of Tb-161 through the irradiation of $^{160}Gd_2O_3$, without compromising its other operational functions. Simulation results indicate that the irradiation of enriched $^{160}Gd_2O_3$ for 14 days yields approximately 4.5 GBq of Tb-161, consistent with the activity required for the purification process to obtain high-activity material for radiolabeling [6,14]. Moreover, the irradiation of larger target masses could enable production at the terabecquerel scale. The RMB complex will also include the necessary infrastructure for target processing and purification of radionuclides such as Tb-161. Furthermore, ongoing research on the enrichment of rare-earth elements via the AVLIS method may provide a reliable source of Gd-160 for large scale production of Tb-161 in the future. Collectively, these features position the RMB as a promising future supplier of Tb-161, should its demand increase, pending its approval for clinical radiotherapy applications.

## ACKNOWLEDGMENTS

We thank the Coaraci Supercomputer for computer time (Fapesp grant #2019/17874-0) and the Center for Computing in Engineering and Sciences at Unicamp (Fapesp grant #2013/08293-7). F.C. Melges, C.G.S. Santos and I.S.R. Júnior thank CNEN/Fundação PATRIA for the research scholarships 011/2024, 006/2025 and 007/2025, respectively, under the FINEP agreements 01.22.0592.00 (RMB172) and 01.24.0373.00 (RMB280).